\documentclass{aastex}
\usepackage{spr-astr-addons}

\RequirePackage{color}

\newcommand{\Ox}{\, ^{16}{\rm O}}
\newcommand{\C}{ \,^{12}{\rm C}}
\newcommand{\Bi}{ \, ^{11}{\rm B}}
\newcommand{\B}{ \, ^{10}{\rm B}}
\newcommand{\Be}{ \, ^{9}{\rm Be}}
\newcommand{\He}{\, ^{4}{\rm He}}
\newcommand{\Hei}{\, ^{3}{\rm He}}
\newcommand{\Hii}{\, ^{2}{\rm H}}
\newcommand{\Hi}{\, ^{1}{\rm H}}
\newcommand{\gbd}{\gamma_{bd}}
\newcommand{\mue}{ \mu_{e}}

\newcommand{\be}{\begin{equation}}
\newcommand{\ee}{\end{equation}} 
\newcommand{\aem}{{ \alpha_{\rm E} }} 
 
\newcommand{\tem}{{ \tau_{\rm E} }} 
\newcommand{\tg}{{ \tau_{\rm G} }} 
\newcommand{\SM}{\rm{M}_{\odot}}
\newcommand{\tG}{t_{\star}}

\begin{document}

\title{The Future Evolution of White Dwarf Stars
Through Baryon Decay and Time Varying Gravitational Constant}


\shorttitle{Long Term Evolution of White Dwarf Stars}
\shortauthors{Ketchum and Adams}

\author{Jacob A. Ketchum\altaffilmark{1}} \and \author{Fred C. Adams\altaffilmark{1,2}}

\altaffiltext{1}{Michigan Center for Theoretical Physics \\
Physics Department, University of Michigan, Ann Arbor, MI 48109}
\altaffiltext{2}{Astronomy Department, University of Michigan, Ann Arbor, MI 48109}

\begin{abstract}
Motivated by the possibility that the fundamental ``constants'' of
nature could vary with time, this paper considers the long term
evolution of white dwarf stars under the combined action of proton
decay and variations in the gravitational constant. White dwarfs are
thus used as a theoretical laboratory to study the effects of possible
time variations, especially their implications for the future history
of the universe.  More specifically, we consider the gravitational
constant $G$ to vary according to the parametric relation $G = G_0 (1
+ t/\tG)^{-p}$, where the time scale $\tG$ is the same order as the
proton lifetime $t_{P}$.  We then study the long term fate and
evolution of white dwarf stars.  This treatment begins when proton
decay dominates the stellar luminosity, and ends when the star becomes
optically thin to its internal radiation.
\end{abstract}


\section{Introduction}

One of the overarching but unresolved questions in physics is whether
or not the fundamental constants are truly constant, or if they could
vary with time. A related question is whether these constants could,
in principle, have other values in far-away regions of space-time
(i.e., in effectively other universes).  Current experiments place
limits on the time variations of the fundamental constants, e.g., the
gravitational constant $G$ and the fine-structure constant $\aem$
(Uzan 2003 and references therein). These constraints can be expressed
in terms of corresponding time scales, $\tg = G / {\dot G}$ and $\tem
= \aem / {\dot \aem}$, which are found in the range $10^{10} -
10^{12}$~yr (Uzan 2003).  Although these time scales are safely longer
than the current age of the universe ($t_0$ = 13.7 Gyr; e.g., Spergel
et al. 2007), these values are much shorter than some time scales that
are experimentally accessible; as one example, the proton lifetime has
a measured lower bound of $10^{33}$ yr (Super-K 1999).  As a result,
over the vast expanses of time available to the universe in the
future, time variations in the physical constants could change the
projected future history of the universe.

This paper addresses this issue by considering possible time
variations of the gravitational constant and its corresponding effects
on the long term evolution of white dwarf stars. In this context, we
are using white dwarfs as a theoretical laboratory to consider the
effects of time-varying $G$. This choice is justified because white
dwarfs are among the simplest and hence most well understood stellar
objects (starting with Chandrasekhar 1939) and because they play an
important role in the future history of the universe (Adams \&
Laughlin 1997, hereafter AL97; see also Cikovic 2003, Dyson 1979,
Islam 1977).  For example, almost all stars turn into white dwarfs
after their nuclear burning phase, and hence a sizable fraction of the
accessible baryonic content of the universe is locked up in white
dwarfs after stellar evolution has run its course (for completeness,
note that a sizable fraction of the baryons remain in the medium
between galaxies in dusters, e.g. Nagamine \& Loeb 2004).

To address this issue, we must define both the time scales and the
functional form of the time variations under consideration.  Although
an enormous variety of time variations are possible, we narrow the
scope by allowing the gravitational constant $G$ to vary according to
the parametric relation
\be
G(t) =  G_0 (1 + t/\tG)^{-p} \, , 
\ee
where $G_0$ is the current value, and where $\tG$ and $p$ are
parameters.  This functional form is perhaps the simplest type of
allowed time variation --- the value of $G$ reduces to the current
value at the current cosmological epoch and decreases as a power-law
in the long term future.  Note that current experimental limits imply
that $\tg = G/{\dot G} > 10^{12}$ yr (e.g., Table IV of Uzan 2003; see
also Barrow 1996) so that $\tG \gg t_0$. In the long term future,
proton decay is one of the most important energy sources for white
dwarfs. This problem thus has two time scales of interest: the proton
lifetime $t_P$ and the time scale $\tG$ for variations in $G$. In the
limit $t_\star \gg t_P$, white dwarf evolution closely follows
previous results obtained using a fixed gravitational constant (AL97,
Adams et al. 1998; Dicus et al. 1982).  As a result, this paper
primarily considers the limit $t_\star \ll t_P$, where white dwarf
evolution is dominated by changes in gravity, and the case where $\tG$
is comparable to $t_P$.

Although this work will focus on time variations in the gravitational
constant $G$, for completeness we note that time variations in other
physical quantities are possible.  For example, the masses of the
fundamental particles or, equivalently, the ratio of electron to
proton mass $\mu = m_e / m_P$, could be time dependent (Calmet \&
Fritzsch 2006).  We note that the quark masses, which play a role in
determining the proton mass, are actually the fundamental constants of
the underlying theory; however, the ratio $\mu$ has been
experimentally constrained (see Uzan 2003 and references therein).

In this paper we consider the variation of $G$ by itself in order to
isolate the effects of its possible time evolution.  However,
fundamental theories (string theory, M theory) suggest that the forces
are unified and hence may vary in strength in a coupled manner (Uzan
2003 provides a brief review of the possibilities).  Unfortunately, we
do not have a definitive working theory of how such coupled time
variations are expected to occur, and we leave such a more comprehensive
treatment for future work.

This paper is organized as follows. In \S 2, we outline a basic model
for white dwarf structure, including our treatment of time variations
in $G$ and proton decay. One interesting complication is that proton
decay leads to a downward cascade for the atomic weights of the
constituent nuclei; this issue is addressed in some detail.  In \S 3
we outline the basic results of the paper, including the H-R diagrams
for the long term evolution of these degenerate stars and their
corresponding chemical evolution. We then conclude in \S 4 with a
summary and discussion of our results.

\section{White Dwarf Model}

\subsection{Conditions and Assumptions}

Given the extremely long time scales associated with baryon decay and
possible time variations of the gravitational constant, we make the
following assumptions about the properties of the white dwarf stars
under consideration here.  The end of conventional star formation is
expected to occur around $\sim 10^{14}$ yr after the big bang (AL97),
and baryon decay will not occur until a cosmic age of $\sim 10^{33}$
yr (Super-K 1999) or later.  As a result, we can safely assume that
the white dwarfs will cool to low temperatures.  Over time scale
$10^{14} - 10^{22}$ yr they will be kept warm through dark matter
capture and annihilation (AL97), but afterwards have only proton decay
as an energy source.  Also, any angular momentum the white dwarfs may
have acquired during formation will be lost, leaving a body free of
rotational deformations.  As a result, electron degeneracy pressure
alone must act to counterbalance gravity and maintain the star's
structure against gravitational collapse.

For white dwarf stars in the low temperature limit, as considered
herein, the physical structure of the star decouples from thermal
considerations.  We can thus assume that the equation of state is that
appropriate for non-relativistic degenerate matter, so that the
pressure is related to the density by the relation
\be \label{eq:eos}
P = K \rho^{5/3} \, , 
\ee 
where the constant $K$ is given by 
\be
K = \frac{\hbar^2}{5 m_e} (3 \pi^2)^{2/3} (m \mue)^{-5/3} \, , 
\ee 
where $m_e$ is the electron mass and $m$ is the atomic mass unit
(Shapiro \& Teukolsky 1983). Here, we take $\mue \equiv \langle A/Z
\rangle $, the average ratio of atomic mass to atomic number for the
entire collection of particles comprising the white dwarf star.  This
ratio will be a function of time due to the effects of baryon decay
(see \S \ref{sec:cascade}.)

\begin{figure}[tb]
\includegraphics[width=\columnwidth]{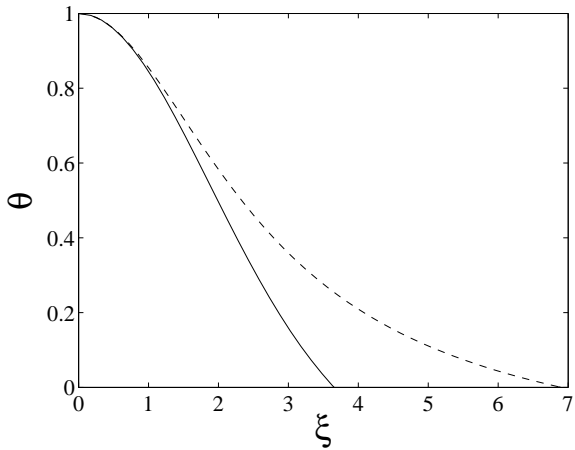}
\caption{Dimensionless form of the density profile given by the
Lane-Emden equation for polytropic index $n=3/2$ (non-relativistic,
\emph{solid line}) and $n=3$ (relativistic, \emph{dashed line}).
Note that $\rho \propto \theta^{n}$.}
 \label{fig:LE}
 \end{figure}

The star must be in hydrostatic equilibrium, which is described by the
Lane-Emden equation, a dimensionless differential equation of the form
\begin{equation}\label{eq:LE}
\frac{1}{\xi^{2}}\frac{d}{d\xi}\xi^{2}\frac{d\theta}{d\xi}= -\theta^{n} \, ,
\end{equation}
where
\[  r = a \xi \, , \]
\[  \rho = \rho_{c} \theta^{n}\, ,  \]
and
\[  a = \left[\frac{(n+1)K\rho_{c}^{(1/n -1)}}{4 \pi G} \right]^{1/2}\, .  \]
In this model, the white dwarf star is a polytropic, self gravitating
gas, with polytropic index $n= 3/2$.  Figure \ref{fig:LE} shows the
density profile of such an idealized white dwarf for the
non-relativistic case ($n= 3/2$).  For comparison, the relativistic
case ($n=3$) is also shown.  Note that white dwarfs are characterized
by $n=3$ (relativistic) polytropes only for masses approaching the
Chandrasekhar mass; the vast majority of white dwarfs produced by the
universe will have lower masses and can be described as $n = 3/2$
polytropes (e.g., Shapiro \& Teukolsky 1983).

The Lane-Emden equation provides a satisfactory description for the
pressure gradient in the interior of the system where the pressure is
extremely high.  However, once the star has radiated away all of its
residual thermal energy, its surface temperature will be nearly
absolute zero.  For the outer layers of a frozen white dwarf, the
typical inter-particle separation is about one Bohr radius, and the
constituent particles naturally fall into a tightly packed lattice
structure.  This densely packed state for the matter in a frozen white
dwarf effectively places a positive, non-zero lower bound on the
density.  Here, we take the lower bound on the density of this lattice
to be a constant value $\rho_{0}~=~1$~g~cm$^{-3}$.  At this density
the coulomb, self-gravitational, and degeneracy energies have
comparable magnitudes and the electrons are no longer degenerate.

Early in a white dwarf's lifetime, this non-degenerate surface layer
is extremely thin, with little of the stellar mass residing therein,
and it is a reasonable approximation to assume that the entire system
is degenerate.  For instance, early in the life of a $1.0\,\SM$ white
dwarf, the non-degenerate 'crust' is only about 1 km deep.  On the
time scales of this paper, as the star expands due to mass loss from
proton decay and varying $G$, it is essential to follow the boundary
between degenerate and non-degenerate layers within the white dwarf.

\subsubsection{Mass}

The total remaining mass of the white dwarf at any time is given by the equation
\begin{equation}\label{eq:totalmass}
M(t) = M_{0} \exp[-\Gamma t]\, ,
\end{equation}
where $\Gamma$ is the baryon decay rate. For the sake of definiteness,
we take $\Gamma = 10^{-37}$ yr$^{-1}$ for this work; the issue of
proton lifetimes is discussed in \S \ref{sec:proton}.

Invoking conservation of mass and the definitions of equation
(\ref{eq:LE}), the mass contribution of the degenerate core enclosed
within the degenerate surface, $\xi_{deg}$, is given by
\begin{equation}\label{eq:degMass}
M_{deg}(t) = \eta_{m} \, \left[ \rho_{c}(t)\,G(t)\,^{-3} \,  \mue(t)^{-5} \right]^{1/2}\, , 
\end{equation}
where
\[
\eta_{m} = \frac{1}{4\sqrt{2\pi}}\left( \frac{\hbar^2}{m_e} 
(3 \pi^2)^{2/3} m^{-5/3}  \right)^{3/2} \xi_{deg}^{2} \, | \theta_{deg}'| \, .
\]  
An alternative way of calculating the total mass is to substitute
$\xi_{1}$ and $|\theta_{1}'(\xi_{1})|$ for $\xi_{deg}$ and
$|\theta_{deg}'|$, respectively, for the expression $\eta_{m}$ in
equation (\ref{eq:degMass}), effectively adding up the mass of a
completely hydrostatic polytropic gas,
\begin{equation}\label{eq:altMass}
\mathcal{M} =\eta_{\mathcal{M}} \frac{M_{deg}}{\eta_{m}} \, ,
\end{equation}
where
\[ 
\eta_{\mathcal{M}} =\eta_{m}
\frac{ \xi_{1}^{2} \, | \theta_{\xi_{1}}'|}{  \xi_{deg}^{2} \, | \theta_{deg}'| }  \, .
\]
The mass contribution $M_{ndeg}$ due to non-degenerate material is
given by the difference between equations (\ref{eq:totalmass}) and
(\ref{eq:degMass}),
\begin{equation}\label{eq:ndegmass}
M_{ndeg} = M-M_{deg}.
\end{equation}
We keep track of the time evolution for all three quantities in our
analysis of white dwarf evolution.  Of particular interest is the
epoch when $M_{deg} = M_{ndeg}$, a time that we define as the end of
the white dwarf's degenerate phase.

\subsubsection{Radius}

The radius of the degenerate core is given by
\begin{equation}\label{eq:Radius}
R_{deg}(t) = \eta_{r}\,\left[ \rho_{c}(t) \, G^{3}(t)\, \mue^{5}(t) \right]^{-1/6}\, ,
\end{equation}
where
\[
\eta_{r} =\frac{1}{2\sqrt{2\pi}} \left( \frac{\hbar^2}{m_e} 
(3 \pi^2)^{2/3} m^{-5/3}  \right)^{1/2}  \xi_{deg} \, .
\]  
Adding the radial contribution of the non-degenerate shell, 
we find an overall white dwarf radius of
\begin{equation}\label{eq:totalrad}
R(t)= \left[ \frac{3 M_{ndeg}}{4 \pi \, \rho_{0}} + R_{deg}^{3}\right]^{1/3} \, .
\end{equation} 
Since the contribution to the overall radius due to non-degenerate
matter is small early on ( only $0.05 \%$ for our adopted lower bound
on density $\rho_{0}~=~1$~g~cm$^{-3}$), an approximate form is
obtained by substituting $\xi_{1}$ for $\xi_{deg}$ in $\eta_{r}$ of
equation (\ref{eq:Radius}).  This approximation effectively assumes
that the full, non-truncated density description is given by the
Lane-Emden equation, so that
\begin{equation}\label{eq:altRad}
\mathcal{R} =\eta_{\mathcal{R}} \frac{R_{deg}}{\eta_{r}} \, ,
\end{equation}
and
\[
\eta_{\mathcal{R}} =\eta_{r}\frac{  \xi_{1}}{\xi_{deg}} \, .
\]
Because we set a lower bound $\rho_{0}$ on the density at the outer
surface, the overall radius $R$ is always less than $\mathcal{R}$,
\,the radius obtained without assuming a minimum density.
Calculations of surface temperature in this work (see \S \ref{sec:TL})
use the radial equation (\ref{eq:totalrad}), whereas previous
calculations (Adams et al. 1998) used the radial form given by
(\ref{eq:altRad}).

\subsubsection{Central Density}

At this point we have expressions for the total mass and total radius
of the white dwarf, where we have included a lower bound $\rho_{0}$ on
the density.  However, both the degenerate mass and radius depend on
the central density of the white dwarf.  The ratio of average density
to central density of a polytropic system is given by
\begin{equation}\label{eq:densi}
\frac{\bar{\rho}}{\rho_{c}}= \frac{3\left|\theta ' (\xi_{1})\right| }{\xi_{1}}\, ,
\end{equation}
where here the average density is described in terms of the
non-truncated mass (\ref{eq:altMass}) and radius (\ref{eq:altRad}),
\emph{i.e.}, 
\begin{equation}\label{eq:avdens}
\bar{\rho} = \frac{\mathcal{M}}{4/3\, \pi \mathcal{R}^{3}}\, .
\end{equation}
Note that this average density represents that corresponding to the
entire profile given by the Lane-Emden equation, not a profile that is
truncated at some non-zero density.  It is also useful in evaluating
equation (\ref{eq:avdens}) to invoke the mass-radius relation
\[
\mathcal{M}\mathcal{R}^{3} = \eta_{\mathcal{R}}^{3} \eta_{\mathcal{M}} G^{-3}\mue^{-5}\,.
\]
Putting these results together, we find that the central density of
the white dwarf can be expressed by the following function of time
alone,
\begin{equation}\label{eq:centdens}
\rho_{c}(t) = \eta_{\rho}\, \left[M^{2}(t)\,G^{3}(t)\,\mue^{5}(t)\right]\,,
\end{equation}
where
\[
\eta_{\rho} = \left( \frac{\xi_{1}}{4 \pi | \theta^{'}(\xi_{1})|}\right) 
\frac{1}{\eta_{\mathcal{R}}^{3} \eta_{\mathcal{M}}} \, \, .
\]
This form provides the correct central density for the white dwarf
while a degenerate core is present (otherwise $\rho_{c} = \rho_{0}$).

\subsubsection{Luminosity and Temperature}\label{sec:TL}

The luminosity of the white dwarf is provided by proton decay and is given by
\begin{equation}\label{eq:lumin}
L(t) = \mathcal{F} c^{2} \Gamma M(t)\, ,
\end{equation}
where $\mathcal{F} \approx 2/3$ is an efficiency factor ($\mathcal{F}
\le 1$ due to neutrino losses, since such particles do not thermalize
within the star).  The surface temperature of the star is given by
\begin{equation}\label{eq:temp}
T^{4}(t) = \frac{ L(t)}{4\pi \sigma_{sb} R^{2}}\,,
\end{equation}
where $ \sigma_{sb} $ is the Stephan-Boltzmann constant, and $R$ is
the radius given by equation (\ref{eq:totalrad}).  Given that $R \le
\mathcal{R}$, the effect of including a lower bound on density is to
provide the star with a somewhat higher temperature, effectively
shifting its trajectory in the HR diagram to the left during late
times.

\subsection{Proton Decay}\label{sec:proton}

Proton decay has many possible channels, but the most relevant
possible processes can be conceptually divided into two types. The
first type of process takes place through unification scale effects,
where the intermediate particle that drives decay is a GUT scale
particle that violates baryon number conservation.  The second type of
process is driven by gravitational effects. In this case, the simplest
description invokes virtual black holes, which also violate baryon
number conservation, as the effective intermediate particles (for
completeness we note that string theories render this description
overly simplistic).  In either case, the estimated time scale for
proton decay is given approximately by the expression 
\be 
t_P = \tau_c \left( \frac{M_X}{m_P} \right)^4 \, , 
\ee 
where $\tau_c \sim 10^{-31}$ yr is the light crossing time of the
proton, $m_P$ is the proton mass, and $M_X$ is the mass of the
intermediate particle that drives the decay process.

For the case of GUT scale proton decay, $M_X \sim M_{GUT} \sim
10^{16}$ GeV; in this case, the proton decay time scale is about
$10^{33}$~yr, roughly the same as the current experimental bounds
(Super-K 1999).  For the case of gravitational proton decay, one
expects $M_X \sim M_{Pl} \sim 10^{19}$~GeV.  This value for the
intermediate particle mass implies a proton decay time scale of about
$10^{45}$~yr, much longer than current bounds. Since the suppression
of baryon number violation at the Planck scale is difficult, we expect
the proton lifetime to fall in the range $10^{33}$~yr $\le t_P \le
10^{45}$~yr. For the sake of definiteness, we take an intermediate 
time-scale between these limits as a working benchmark value so that 
$\Gamma^{-1} \sim t_P \sim 10^{37}$~yr.

For the case of gravitationally driven proton decay with time varying $G$,
an additional complication arises. The Planck mass, which sets the
effective scale for the virtual black holes, scales as $M_{Pl} \propto
G^{-1/2}$, so the (gravitationally driven) proton decay time scales as
$t_P \propto G^{-2}$. As a result, the proton lifetime is expected to
increase as the gravitational constant decreases. 

\subsection{Chemical Cascade Chain} \label{sec:cascade}

This section investigates the variations of $\mue \equiv \langle A/Z
\rangle$, which determines, in part, the equation of state.  This
quantity changes with time due to nuclear decay events that determine
the long term fate of white dwarf stars.  Initially, a white dwarf
star will most likely consist of a mixture of heavier elements such as
$\He, \C$, and $\Ox$.  On sufficiently long time scales, however,
random baryon decay events send these particles on a cascading path
towards $\Hi$, and eventually total extinction.  To first order,
$\mue$ is proportional to the mean value $\langle A/Z \rangle$ of the
white dwarf's constituent elements.  This ratio will be nearly 2 for
any white dwarf consisting mostly of heavier elements.  By the time
that about half of the original nucleons have decayed through proton
decay processes, however, a significant portion of the composite mass
will be in the form of Hydrogen, for which $\mue = 1$.  We thus expect
$\mue$ to be a decreasing function of time, with $\mue \rightarrow 1$
in the long time limit $t \rightarrow \infty$.

This cascade of particles is categorized below.  This chain begins
with $\C$, and continues down to baryonic extinction (Thornton \& Rex
2000); note that we could start with larger nuclei (e.g. Oxygen), but
the latter parts of the chain would remain the same.  The basic
reactions include:
\be
\C - b \longrightarrow \Bi + \gbd
\ee
\be
\Bi -b \longrightarrow \B + \gbd
\ee
\be
\begin{aligned}
\B 
	\begin{cases}
	   - \,p^{+} &\longrightarrow \Be +  \gbd\\
	   -\, n &\longrightarrow 2 \He+ \Hi + \gbd
	\end{cases}
\end{aligned}
\ee
\be
\Be - b \longrightarrow 2 \He + \gbd
\ee
\be
\He - b \longrightarrow \Hei + \gbd
\ee
\be
\begin{aligned}
\Hei 
	\begin{cases}
	  - \,p^{+} &\longrightarrow \Hii + \gbd\\
	  - \,n &\longrightarrow 2 \Hi + \gbd
	\end{cases}
\end{aligned}
\ee
\be
\Hii - b \longrightarrow \Hi +\gbd
\ee
\be
\Hi - p^{+} \longrightarrow \gbd
\ee

These reaction equations show the nuclei resulting from each
particular baryon decay event and an energy release.  Each reaction
given above may actually include several different reactions en route
to the production of the listed stable particles.  Among the decay
products will be some particle with a positive charge, often a
positron, which annihilates with an electron and turns into energy
represented by the photons $\gbd$ in the reaction equations.  Among
the baryon decay events listed above, $(b)$ signifies either a proton
or a neutron decaying, $(n)$ signifies neutron decay, and $(p^{+})$
signifies proton decay.  The particular daughter particles for $\B$
and $\Hei$ are dependent upon whether a proton or a neutron decays,
whereas the other particles listed in the chain produce only one
variety of products, regardless of whether a proton or a neutron
decays.  Although most reactions listed above have outcomes that are
independent of which particle decays, the exact path taken to achieve
the stable particles could vary (and are not documented in the chain).

\begin{figure}[t]
	\includegraphics[width=\columnwidth]{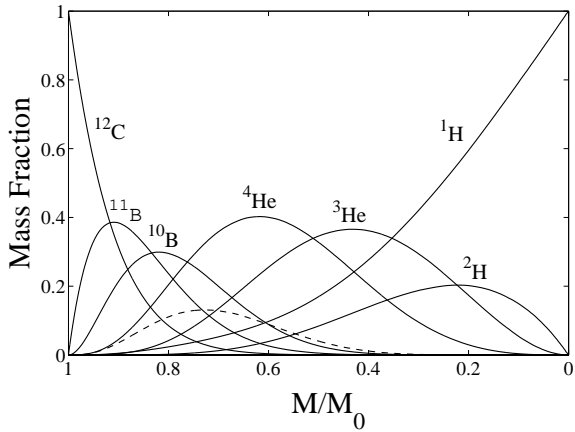}
	\caption{Chemical species mass fraction versus relative mass
          of white dwarf with initial composition of pure $\C$.  Each
          curve represents a species of particle within the white
          dwarf during proton decay.  The labels indicate the particle
          species, with the exception of $^9$Be, which is represented
          by the dashed curve.}\label{fig:species}
\end{figure}

To model this process, we used a Monte Carlo simulation to follow the
evolution of the particle species population as a function of overall
mass for a white dwarf star undergoing baryonic decay.  In these
simulations, both protons and neutrons were allowed to decay with
equal likelihood; since proton decay (neutron decay in bound nuclei)
has not yet been measured, this assumption remains unverified
experimentally. If the resultant nucleus is unstable to decay, then
the decay was immediately allowed to take place (since the proton
decay time-scale is far greater than the nuclear decay time-scale of
any known isotopes of light elements.)  Figure~\ref{fig:species} shows
the results of one such Monte Carlo simulation for an initial
collection of $\sim 10^{6}$ atoms of $\C$, which were subject to
baryon decay (see also AL97).  The implications of this chemical
cascade on the values of $\mue$, and hence the equation of state, are
discussed below.

\begin{figure}[t]
	\includegraphics[width=\columnwidth]{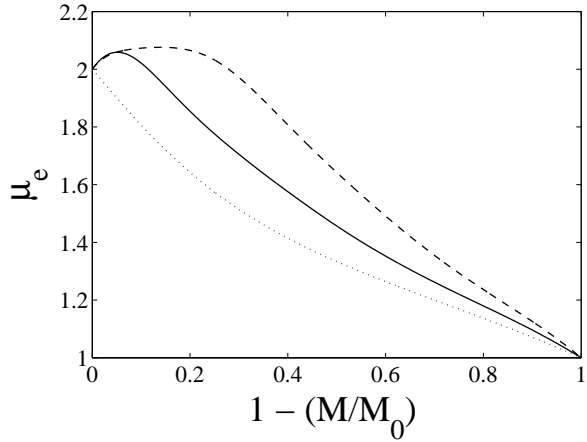}
	\caption{The ratio $\mue$ as a function of decayed mass
          fraction for white dwarf compositions of initially pure
          $^{16}0$ (\emph{dashed}), $\C$ (\emph{solid}), and $\He$
          (\emph{dashed}).}\label{fig:AZ}
\end{figure}

For completeness, we note that the values of $\mue$ can be affected by
spallation and pycnonuclear reactions.  In particular, pycnonuclear
reactions work to build up the abundance of $\He$ from smaller
elements, effectively widening and raising the $\He$ curve of Figure
\ref{fig:species} at the expense of the $\Hei$, $\Hii$, and $\Hi$
curves.  The effect is to increase the value of $\mue$ and hence the
height of the curve in Figure \ref{fig:AZ} near the low mass end.
Spallation events slowly redistribute protons and neutrons one at a
time among the nuclei in the star.  This redistribution happens at a
rate that is approximately one per baryon decay event.  Thus,
spallation has a similar effect as pycnonuclear reactions in that it
acts to increase the size of the nuclei, albeit at a slow rate, and
slightly increases the value of $\mue$ near the end of the white
dwarf's life.

Figure 3 shows the ratio $\mu_e$ as a function of time, where time is
given in terms of the white dwarf mass. These curves are calculated
from the Monte Carlo simulations described earlier.  Initially $\mue =
2$ for all three curves, corresponding to their respective homogenous
compositions.  For a brief period for the $\C$ curve, this ratio rises
above 2 to an approximate value of $2.06$ with nearly 95\% of the
initial mass remaining.  This ratio then decreases monotonically
towards a value of 1 (homogeneous $\Hi$ composition).  The $\Ox$ curve
has a much broader hump above the $\mue = 2$ line and reaches a
maximum value of approximately $2.07$ with nearly $86\%$ of the
initial mass remaining. These curves would rise slightly between the
end points by including spallation and pycnonuclear effects in the
Monte Carlo simulation.

For simplicity, we use a fitting function that represents the $\mue$
curves of Figure \ref{fig:AZ} with an accuracy of $\sim 1\%$. The
fitting function has the form
\begin{equation}\label{eq:formfit}
\mue(u) = 2 - u + \left(a_{1}u + a_{2}u^{2} + a_{3}u^{3}\right)\, \exp (a_{4}u)\, ,
\end{equation}
where $u = 1 - M/M_{0}$.  Although one can find a good fit to these
curves, there exists no unique set of fitting parameters $(a_{1},
a_{2}, a_{3}, a_{4})$.  Clearly, $a_{1}+a_{2}+a_{3}=0$ in order to
fulfill the requirement that $\mue(1) = 1$, .  But, more specifically,
for a given $a_{4}$, there exists no unique pair $(a_{1}, a_{2})$ that
provides a fit to within a given accuracy.  In Table \ref{T:param} we
provide suggested fitting parameters for three different cases where
the initial white dwarf composition is taken to be of pure $\He$,
$\C$, or $\Ox$ (note that we give a large number of significant
figures in this table).

\begin{table}[h!]
\caption{Fitting parameters for $\mue$ time evolution curves.  The
  Monte Carlo simulations were started with a pure concentration of
  the chemical species listed in the left column.  The time evolutions
  of $\mue$ are recovered by substituting the parameters from this
  table into equation (\ref{eq:formfit}).}\label{T:param}
\begin{tabular}{@{}rcccc}
\hline
Species  &  $a_{1}$ & $a_{2}$ & $a_{3}$ & $a_{4}$\\
\hline
$\He$  & -0.952 & 3.23 & 4.182 & -3.025 \\
$\C$  & 3.1218 & -12.9250 & 9.24125 & -4.95509 \\
$\Ox$  & 1.8935 & 14.3400& -16.2335 & -5.6545 \\
\hline
\end{tabular}
\end{table}

\section{Results} \label{sec:results}  

Given the model for white dwarf structure constructed in the
previous section, we consider the long term evolution of these
degenerate stars through the action of proton decay and time varying
gravitational constant. To start, we first consider the case where
gravity varies but no proton decay takes place. Under these
conditions, the white dwarf grows in radial size with time. Both the
radius of the degenerate interior and the fraction of the star that
makes up the non-degenerate outer layer are increasing functions of
time.

\begin{figure}[tb]
	\includegraphics[width=\columnwidth]{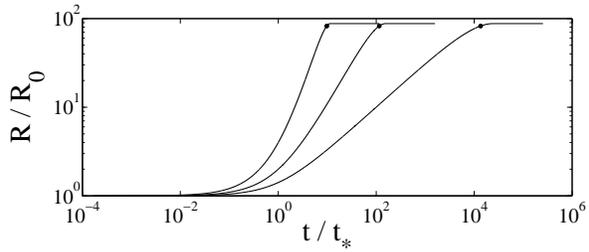}
	\caption{ Radius versus time diagram for white dwarf evolution
          with constant $\C$ composition (\emph{no baryon decay}) and
          time varying G with $t_{\star} = 10^{37}$ yr; $ p = 0.5$
          (\emph{right}), $1.0$ (\emph{middle}), and $2.0$
          (\emph{left}).  The mass is $1 \SM$.  Radius and time are
          shown in dimensionless form.  The dots mark the end of the
          white dwarf's degenerate phase.  In all three cases the
          white dwarf grows to almost 100 times its original radius,
          but the time scale on which this growth occurs varies with
          $p$.}\label{F:rvtnopd}
\end{figure}

The overall radius increases for two reasons: (1) reducing gravity's
strength effectively reduces the central density of the white dwarf,
allowing the degenerate core to spread out, and (2) the reduction of
the degenerate core mass results in an increase in mass of the frozen
solid lattice (approximately at constant density), leading to an
increase in its radius.  If gravity weakens enough, the pressure
exerted by gravity at the center of the white dwarf decreases below
the $\sim10^{12}$ dyne cm$^{-2}$ lower bound, completely dissolving
the degenerate core into the outer solid lattice.  In the long term,
the white dwarf is left with a radial density that is practically
uniform and a maximal surface radius.  Figure \ref{F:rvtnopd} shows
the radius as a function of time for this scenario with different
rates of gravitational time scales $\tG$ for a $1 \SM$, white dwarf
with pure $\C$ composition.  This figure shows that under
gravitational time variations, the white dwarf grows to nearly 100
times its original radius before becoming a frozen lattice composed
fully of non-degenerate matter.

\begin{figure}[tb]
	\includegraphics[width=\columnwidth]{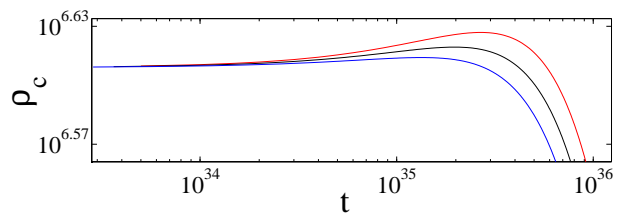}
	\caption{ Central density versus time for white dwarf
          evolution with initial $\C$ composition and
          $\Gamma^{-1}=t_{\star} = 10^{37}$ yr; $ p = 0.0$
          (\emph{red}), $0.25$ (\emph{black}), and $0.5$
          (\emph{blue}).  The starting mass is $1 \SM$.  Notice that
          the central density increases before falling off rapidly for
          all three cases.  This feature of the central density is due
          to the initial increase in $\mue$ for $\C$ and thus is
          expected to occur for an $\Ox$ white dwarf as well.  Density
          is shown in g~cm$^{-3}$ and time is given in
          years.}\label{fig:rhovtC}
\end{figure}

\begin{figure}[tb]
	\includegraphics[width=\columnwidth]{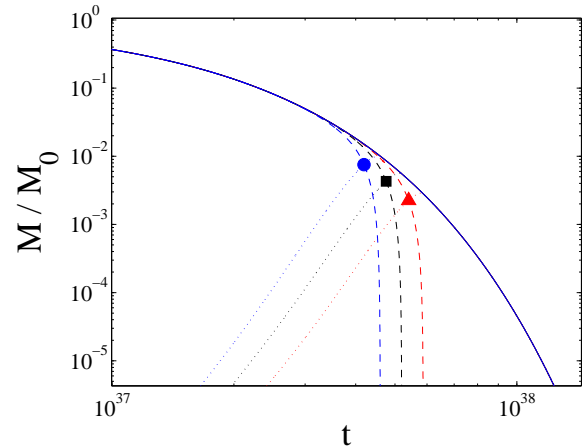}
	\caption{Mass versus time for white dwarf evolution through
          proton decay with an initial $\C$ composition and $t_{\star}
          = \Gamma^{-1} = 10^{37}$ yr; $ p = 0$ (\emph{red}), $0.25$
          (\emph{black}), and $0.5$ (\emph{blue}); $M(t)$
          (\emph{solid}), $M_{deg}$ (\emph{dashed}), $M_{ndeg}$
          (\emph{dotted}).  Mass is shown as the remaining fraction
          and time is shown in years.  The triangle, square, and
          circle mark the fractional mass and time when the mass in
          non-degenerate form equals the amount of degenerate mass for
          gravitational index $p = 0$, $0.25$, and $0.5$,
          respectively.  This crossover point defines the time when
          the white dwarf's degenerate phase ends and its
          non-degenerate phase begins. }\label{fig:MvstC}
\end{figure}

\begin{figure}[tb]
	\includegraphics[width=\columnwidth]{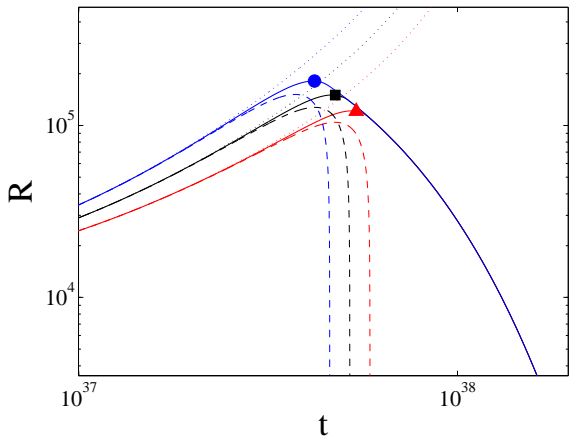}
	\caption{Radius versus time for white dwarf evolution through
          proton decay with initial $\C$ composition and $t_{\star} =
          \Gamma^{-1} = 10^{37}$ yr; $ p = 0$ (\emph{red}), $0.25$
          (\emph{black}), and $0.5$ (\emph{blue}); $\mathcal{R}$
          (\emph{dotted}), $R$ (\emph{solid}), $R_{deg}$
          (\emph{dashed}).  Radius is shown in km and time is shown in
          years.  The triangle, square, and circle correspond to those
          shapes in Figure \ref{fig:MvstC}.  In all cases, these
          degeneracy transitions occur shortly after the white dwarf
          has reached maximum radius and has began to
          shrink.}\label{fig:RvstC}
\end{figure}

\begin{figure}[tb]
	\includegraphics[width=\columnwidth]{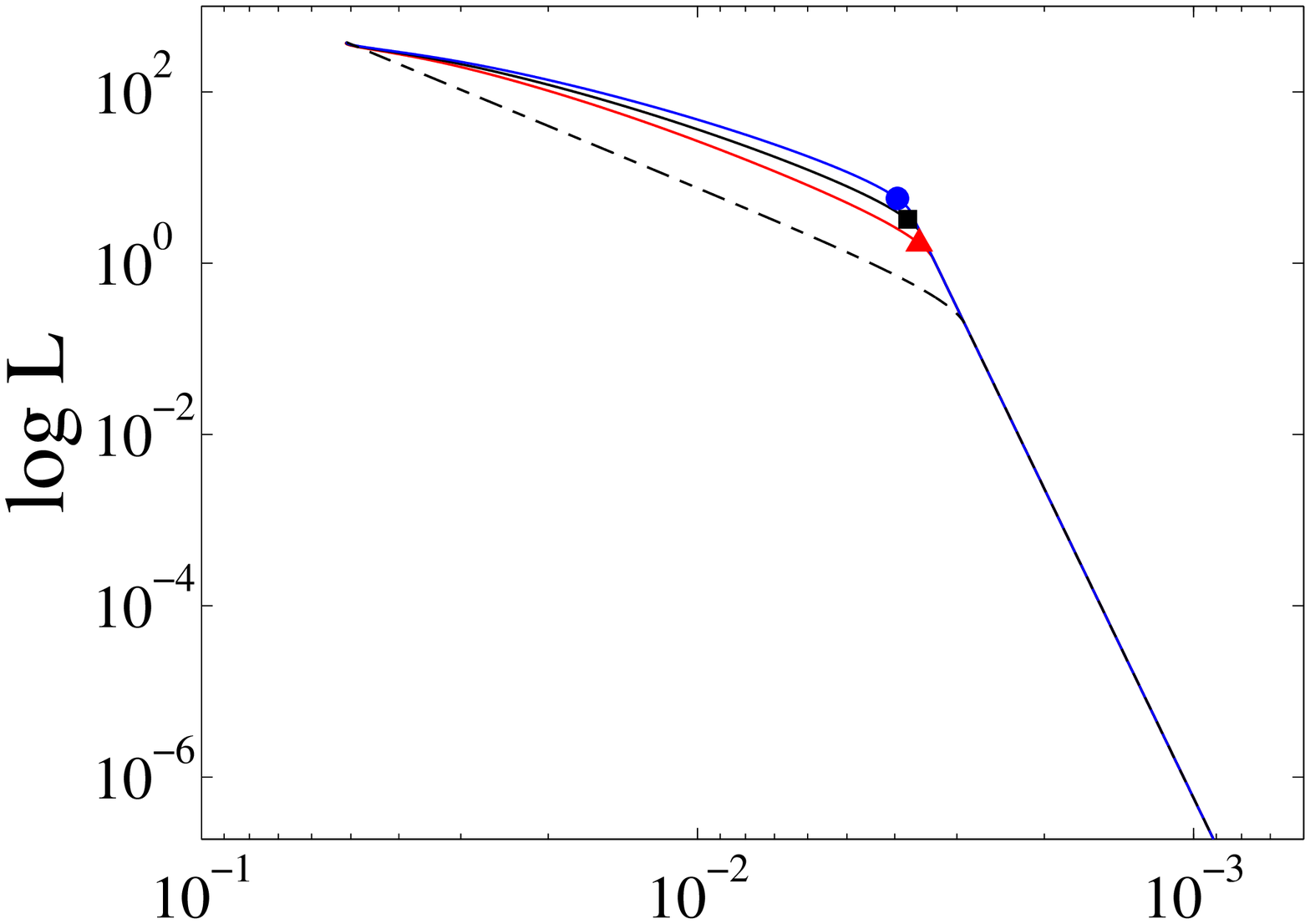}
	\includegraphics[width=\columnwidth]{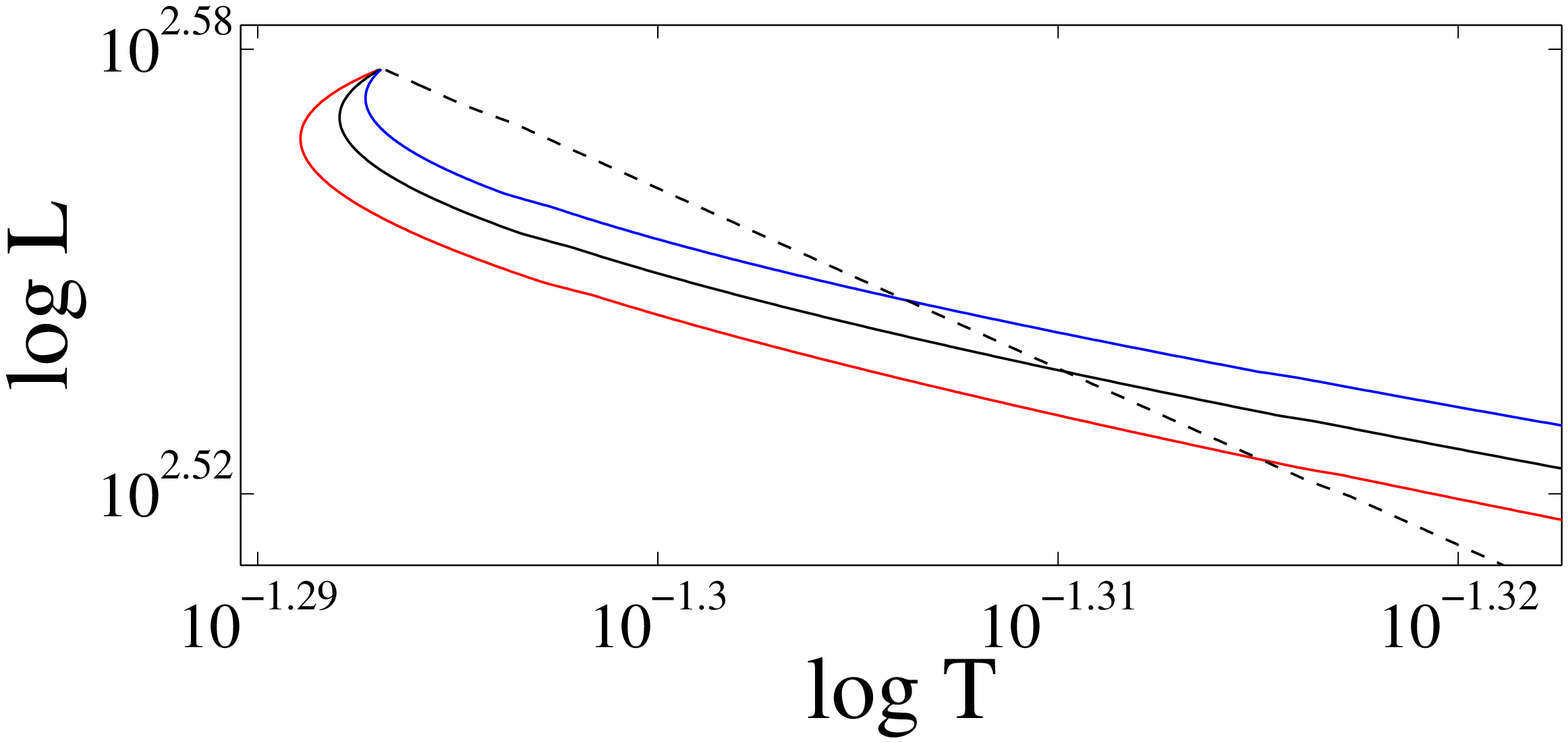}
	\caption{H-R Diagram for white dwarf evolution including
          proton decay and time variations in the gravitational
          constant.  In all cases, the star begins with mass $M$ = 1.0
          $\SM$, a pure $\C$ composition, and evolves according to
          $\tG = \Gamma^{-1}$ = $10^{37}$ yr.  The three curves show
          different values of the gravitational index $p$ = 0 (red),
          0.25 (black), and 0.5 (blue). Temperature is shown in Kelvin
          and luminosity is shown in Watts.  The triangle, square, and
          circle correspond to those shapes in Figure \ref{fig:MvstC}.
          The dashed line shows the reference case of a white dwarf
          with constant $\mue = 2$ and no variations in the
          gravitational constant $(p=0)$. The lower panel shows a
          close-up of the beginning of the time evolution, for the
          first $\sim10 \%$ of the mass loss. During this early phase,
          $\mue > 2$, the radius shrinks slightly, and the
          photospheric temperature heats up.}\label{fig:HRC}
\end{figure}

We also consider the case of constant gravity while subjecting the
star to baryon decay, which causes the star to lose mass.  The white
dwarf increases in radial size until its mass in degenerate matter
decreases to only about two thirds of the total stellar mass.  After
this point, the radius and mass of the white dwarf shrink until the
internal coulomb forces dominate over gravity, and the object
resembles a rock. This scenario is captured by the red curves in
Figures \ref{fig:rhovtC}, \ref{fig:MvstC}, \ref{fig:RvstC}, and
\ref{fig:HRC}.

Next we consider the effects of both time variations in the
gravitational constant and baryon decay. The interplay between these
two effects will be the focus of the remainder of this section.  As
shown below, the main result is to reduce the ratio of degenerate to
non-degenerate matter within the star at the time when the remnant
begins its shrinking phase.

Figures \ref{fig:MvstC} and \ref{fig:RvstC} show the time behavior of
mass and radius, respectively, for a $1.0$ M$_{\odot}$ white dwarf
composed initially of pure $\C$, subjected to baryonic decay with
characteristic time scale of $\Gamma = 10^{-37}$ yr$^{-1}$.  The three
different sets of curves represent various indices of gravitational
weakening with time constant $\tG = 10^{37}$ yr.  In Figure
\ref{fig:MvstC}, one can see that the overall mass envelope
(\emph{solid line}) for a white dwarf under these conditions is
independent of the gravitational index $p$ and is fully described by
$\Gamma$, the characteristic time for baryon decay.  The set of red
curves show the case of a static gravitational constant ($p = 0$).  As
baryons decay, more material precipitates out of the degenerate core
(\emph{dashed line}) to join the outer, non-degenerate layer
(\emph{dotted line}).  At time $t \approx 5.4 \times 10^{37}$ yr
($\Gamma t = 5.4$), nearly 99.5\% of the white dwarf's initial mass
has decayed and the mass in the degenerate core is less than the mass
in the non-degenerate outer lattice.  This event is marked by the
solid, red triangles in Figures \ref{fig:MvstC}, \ref{fig:RvstC}, and
\ref{fig:HRC}, and marks the time when the degenerate phase of the
star comes to an end.  This event does not spell the end of the white
dwarf's life, which goes well beyond $t \approx 5.9 \times 10^{37}$ yr
($\Gamma t = 5.9$) when the degenerate core disappears entirely from
the stellar remnant. As shown in Figure \ref{fig:RvstC}, this time
also approximately marks the point when the star begins to shrink in
radius.  At this epoch, $R \approx 1.2 \times 10^{5}$ km, a size that
is comparable to that of Uranus or Neptune.

Figure \ref{fig:HRC} shows the H-R diagram, which summarizes the
evolution of these white dwarf stars.  While the star remains (mostly)
degenerate, the tracks in the H-R diagram are approximately given by
$L \sim T^{12/5}$. This form would be exact for an object that obeys
the usual white dwarf (degenerate) mass-radius relation and has
luminosity proportional to its mass.  Moderate deviations from this
simple behavior occur due to variations in the chemical composition,
which affect the equation of state (and due to time variations in $G$
-- see below).  After the degenerate phase ends, the tracks in the H-R
diagram become much steeper with $L \sim T^{12}$.  This form applies
for a constant density object with its luminosity proportional to its
mass.  At this transition point, 99.5\% of the mass has decayed away,
the chemical composition for this white dwarf is $\sim 99\% \Hi$,
$\sim 1 \% \Hii$, with traces of $\Hei$ and $\He$ (see Figure
\ref{fig:AZ}).  The remnant now amounts to a medium sized planet,
composed mostly of Hydrogen ice, glowing extremely dim with an output
of only $\sim 2$ Watts (Figure \ref{fig:HRC}).  This huge ball of
Hydrogen ice fizzles away as the white dwarf continues to fade and
shrink.  Finally, around $2.2 \times 10^{38}$ yr ($\Gamma t = 22$),
when the mass is $\sim 10^{21}$ kg, radius $R = 2.9 \times 10^{2}$ km,
effective temperature is $T = 1.3 \times 10^{-3}$ K, and luminosity $L
=1.9 \times 10^{-7}$ Watts, the remnant ceases to be a star.

This story changes only slightly with the introduction of a time
varying gravitational constant.  The set of black curves in these
Figures shows what happens when the gravitational index is $p = 0.25$,
and blue curves represent a gravitational index $p = 0.5$.  For the
scenario where $p = 0.25$, the degenerate phase ends at $t \approx 4.8
\times 10^{37}$ yr ($\Gamma t = 4.8$) when 99.1\% of the mass has
decayed away.  At this time, the star is about $1.5 \times 10^{5}$ km
in radius and has a power output of about 3.3 Watts.  This white dwarf
is $\sim 98.2\% \Hi$, $\sim 1.8 \% \Hii$, and $\sim 0.07 \% \Hei$ by
mass, and is thus nearly identical chemically to the white dwarf
evolving with a static gravitational constant.  For the scenario where
$p = 0.5$, the degenerate phase ends at $t \approx 4.2 \times 10^{37}$
yr ($\Gamma t= 4.2$) when 98.5\% of the mass has decayed away.  At
this time, the star is about $1.8 \times 10^{5}$ km in radius and has
a power output of about 5.7 Watts.  This white dwarf is $\sim 97.1\%
\Hi$, $\sim 2.7 \% \Hii$, and $\sim 0.1 \% \Hei$ by mass, and is also
nearly chemically identical to the white dwarf produced using a fixed
gravitational constant.

The white dwarf completes its transformation into a rock of hydrogen
ice when it becomes optically thin to the radiation that originates
from within.  This condition is determined by the star's column
density.  The star contains two types of radiation that must be
considered when determining this transition.  The first type is the
$\sim 235$ MeV gamma rays produced in each baryon decay event; the
stellar remnant becomes optically thin to this radiation at an
approximate radius of 1 meter (for our assumed $\rho_{0}$).  The other
type is the radiation that comes from the thermalization of these
gamma rays, which produce photons with wavelength $\lambda \sim
10^{2}$ cm.  The white dwarf becomes optically thin to this radiation
at a mass of about $10^{21}$ kg and radius of $2.9 \times 10^{2}$ km.
This mass is calculated by using the average density of the white
dwarf, a value that depends only upon coulomb forces between
neighboring particles in the Hydrogen lattice and not upon gravity.
As a result, the transition from white dwarf to rock for the two
scenarios with $p = 0.25$ and $p = 0.50$ depends only upon $\Gamma$,
and thus happens at the same time, mass, radius, temperature, and
luminosity as the case with $p = 0$.  Corrections to this
approximation could come from accounting for the exact chemical make
up and crystal grain structure of the solid lattice, both of which
would differ slightly for the three indices.
  
\begin{figure}[tb]
	\includegraphics[width=\columnwidth]{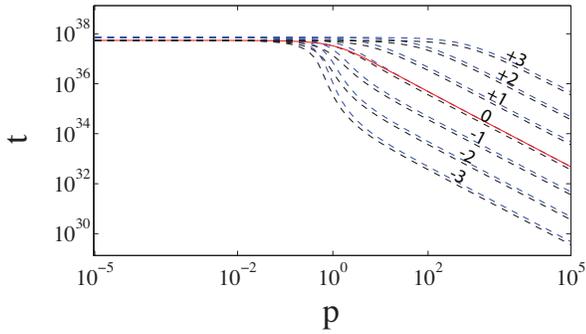}
	\caption{Time duration of the degenerate phase of white dwarf
          evolution as a function of the gravitational index $p$.  The
          proton decay time parameter $\Gamma = 10^{-37}$ yr$^{-1}$and
          the gravitational strength time parameter $t_{\star} =
          10^{37}$ yr.  The initial mass of the white dwarf is $1\,
          \SM$.  Each set of curves contains one black ($\mue = 1$)
          and one blue ($\mue = 2$) dashed line and are labeled by an
          integer $\ell = -3,-2,...,3$ where $\ell = \log(\Gamma\,
          \tG)$.  The time marking the end of the degenerate phase is
          largely independent of $p$ when $\tG \ge \Gamma^{-1}$ and
          for $p < 1$.  For values of $p \gg 1$, the different curves
          become monotonically decreasing, parallel lines.  The solid
          curve represents a white dwarf star initially composed of
          pure $\C$.}\label{fig:TvsP}
\end{figure}

\begin{figure}[tb]
	\includegraphics[width=\columnwidth]{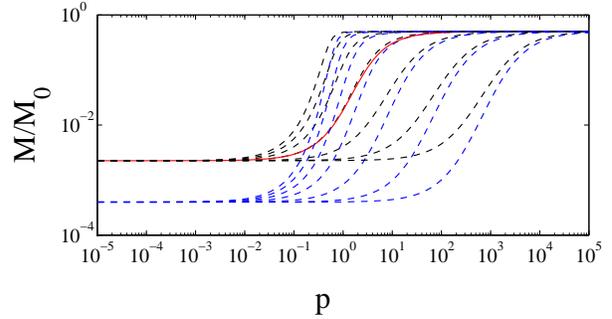}
	\caption{Mass fraction in degenerate matter versus
          gravitational index $p$.  Each set of curves contains one
          black ($\mue = 1$) and one blue ($\mue = 2$) dashed line and
          correspond to an integer $\ell = -3,-2,...,3$, from left to
          right, where $\ell = \log(\Gamma\, \tG)$.  The dynamical
          range for $\Gamma = t_{\star}$ has $p$ spanning 4 orders of
          magnitude centered at $p=1$.  The solid curve represents a
          white dwarf star initially composed of pure
          $\C$.}\label{fig:MvsP}
\end{figure}

Figures \ref{fig:TvsP} and \ref{fig:MvsP} show curves characterizing
the time and mass marking the end of the degenerate phase of the white
dwarf for a continuum of gravitational indices spanning 10 orders of
magnitude, and seven characteristic time scales for gravitational time
variations.  These graphs show the limiting behavior for $p \ll 1$ and
$p \gg 1$ and for intermediate values.  For instance, in the benchmark
case of a pure $\C$ white dwarf, the dynamical range of index
parameter is $0.01 \le p \le 100$.  If $p$ takes on values lower than
this range, the fate of the white dwarf is dominated by proton decay;
if $p$ takes on values larger than this range, then the fate of the
white dwarf is dominated by gravitational time variations.  This
dynamical range of index parameters broadens and increases as the
product $\Gamma\tG$ increases, which indicates that gravity must vary
rapidly to compete with a proton decay.

Before leaving this section, we briefly consider the issue of the
``starting'' mass.  First, note that the ``starting'' mass in this
context is the ``final'' mass in stellar evolution calculations.  At
the end of their nuclear burning phase, stars become white dwarfs with
masses in the range $0.08 \SM \le M \le 1.4 \SM$. To leading order,
white dwarfs with larger masses must become white dwarfs with smaller
masses as their constituent baryons decay.  As a result, white dwarf
evolution should not depend sensitively on the starting mass (in the
long term). However, the chemical composition plays a role in
determining stellar structure, primarily through the value of $\mue$,
which in turn determines the equation of state.  White dwarfs with
larger initial masses must lose a larger percentage of their mass
before becoming non-degenerate, and hence will experience a greater
change in chemical composition. One might expect the stellar
properties at the transition point (from degenerate to non-degenerate
stars) to depend on the starting stellar mass. However, the larger
white dwarfs begin their evolution with larger nuclei (e.g., Carbon
and Oxygen), whereas small white dwarfs (a few 0.1 $\SM$) are made up
mostly of Helium. These two effects largely compensate, so that white
dwarfs of all starting masses become non-degenerate with roughly the
same properties.

To illustrate this point, consider white dwarfs subject to both baryon
decay and time varying $G$, with the standard time scales $\tG =
\Gamma^{-1}$ = $10^{37}$ yr and with $p$ = 1. We can then examine
stellar properties at the time when the stars become non-degenerate.
For a small 0.1 $\SM $ white dwarf initially composed of $\He$, this
transition occurs at $t = 1.8\times10^{37}$ yr ($\Gamma t = 1.8$)when
$\mue \approx 1.11$, the mass $M=1.6\times10^{-2}\SM$, and the radius
$R=1.9\times10^{6}$ km.  For comparison, for a larger 1.0~$\SM$ white
dwarf initially composed of $\C$, this transition occurs at $t = 3.3
\times 10^{37}$ yr ($\Gamma t = 3.3$) when $\mue \approx 1.03$, the
mass $M=3.8\times10^{-2}\SM$, and the radius $R=2.45\times10^6$~km.
These differences are thus at the factor of $\sim 2$ level.

\section{Conclusion}

This paper has constructed a model for the structure and evolution of
white dwarf stars under the action of both proton decay and time
variations in the gravitational constant (section 2). This model
includes variations in chemical composition (Figure 2 and section 2.3)
which affects the equation of state.  The model also includes separate
accounting for the inner degenerate regions (which obey the polytropic
equation of state [1]) and the outer non-degenerate regions.

Our main findings can be summarized as follows: 

In the absence of baryon decay, white dwarfs grow larger as the
strength of gravity decreases (Figure 4). The radius reaches a maximum
value when the entire star becomes non-degenerate; at this time, the
radius is $\sim 100$ times larger than its starting value.

With the inclusion of baryon decay, white dwarfs lose mass with time
and follow tracks in the H-R diagram as shown in Figure
\ref{fig:HRC}. In all cases, after an initial transient phase, the
stars grow dimmer and redder with time, and hence move to the lower
right in the H-R diagram.  While the stars remain (primarily)
degenerate, the tracks have a shallow slope, with $L \propto
T^{12/5}$.  After the stars lose enough mass to become primarily
non-degenerate, the tracks become much steeper, with $L \propto
T^{12}$. These slopes would be exact in the absence of changes in the
chemical composition; the chemical variations and the time variations
in $G$ lead to small departures from this behavior.

During the degenerate regime of evolution, the radii of these white
dwarf stars grow larger due to mass loss from proton decay and due to
the weakening of gravity. During the later, non-degenerate phase, the
radii become smaller. The maximum value of the radius occurs near the
transition between the degenerate and non-degenerate regimes, with a
value that is typically $\sim 10$ times the starting radius (see
Figure \ref{fig:RvstC}).

In addition to understanding the long term fate and evolution of white
dwarfs, one motivation for this work is to understand the larger issue
of the future of the universe.  Previous projections of the future
(e.g. AL97, Dyson 1979) are predicated on the assumption that the laws
of physics are known and will not change with time.  One important
issue is thus the question of whether or not the constants of nature
change with time, and how such variations would change projections of
our future history.  In the case of white dwarf evolution, we find
that time variation leads to quantitative --- but not qualitative ---
changes in the future timeline.  Time varying $G$ leads to variations
in the exact mass, size, and composition of white dwarfs as they
become non-degenerate rock-like objects.  However, the starting states
and final states are essentially the same, so time variations in
gravity do not lead to fundamental changes in the overall picture.
White dwarfs start as degenerate objects with the current value of
$G$, and hence are constrained to begin their evolution in nearly the
same states.  At the other end of time, white dwarfs cease to act as
stars when they become optically thin to their internal radiation.  At
this epoch, the ``stars'' are essentially large rocks of Hydrogen ice,
and their structure is non-degenerate and largely independent of
gravity.  As a result, time variations in $G$ primarily affect the
intermediate states, and this work shows that the modifications are
relatively modest (see Figures \ref{fig:rhovtC} - \ref{fig:HRC}).

\acknowledgments
We thank Jeff Druce for useful discussions. This work
was supported by the Foundational Questions Institute through Grant
RFP1-06-1 and by the Michigan Center for Theoretical Physics.

\nocite{*}
\bibliographystyle{spr-mp-nameyear-cnd}
\bibliography{biblio-u1}

\end{document}